\documentclass[conference]{IEEEtran}
\IEEEoverridecommandlockouts

\usepackage{cite}
\usepackage{amsmath,amssymb,amsfonts}
\usepackage{algorithmic}
\usepackage{graphicx}
\usepackage{textcomp}
\usepackage{xcolor}
\usepackage{graphicx}
\usepackage{booktabs, tabularx}
\usepackage{comment}
\usepackage{hyperref}

\makeatletter
\newcommand{\linebreakand}{%
  \end{@IEEEauthorhalign}
  \hfill\mbox{}\par
  \mbox{}\hfill\begin{@IEEEauthorhalign}
}
\makeatother

\def\BibTeX{{\rm B\kern-.05em{\sc i\kern-.025em b}\kern-.08em
    T\kern-.1667em\lower.7ex\hbox{E}\kern-.125emX}}
\begin{document}

\title{MC-SEMamba: A Simple Multi-channel Extension of SEMamba 
}

\author{\IEEEauthorblockN{Wen-Yuan~Ting}
\IEEEauthorblockA{
\textit{Academia Sinica}\\
Taipei, Taiwan \\
dingwenyuan22@citi.sinica.edu.tw}\\
\and
\IEEEauthorblockN{Wenze Ren}
\IEEEauthorblockA{
\textit{National Taiwan University}\\
Taipei, Taiwan  \\
r11942166@ntu.edu.tw}\\
\and
\IEEEauthorblockN{Rong Chao}
\IEEEauthorblockA{
\textit{National Taiwan University}\\
Taipei, Taiwan   \\
roychao@cmlab.csie.ntu.edu.tw}\\
\linebreakand
\IEEEauthorblockN{Hsin-Yi~Lin}
\IEEEauthorblockA{
\textit{Seton Hall University}\\
South Orange, NJ, USA  \\
hsinyi.lin@shu.edu}\\
\and
\IEEEauthorblockN{Yu~Tsao}
\IEEEauthorblockA{
\textit{Academia Sinica}\\
Taipei, Taiwan \\
yu.tsao@citi.sinica.edu.tw}\\
\and
\IEEEauthorblockN{Fan-Gang Zeng}
\IEEEauthorblockA{
\textit{University of California Irvine}\\
Irvine, CA, USA \\
fzeng@uci.edu}
}

\maketitle

\begin{abstract}
Transformer-based models have become increasingly popular and have impacted speech-processing research owing to their exceptional performance in sequence modeling. Recently, a promising model architecture, Mamba, has emerged as a potential alternative to transformer-based models because of its efficient modeling of long sequences. In particular, models like SEMamba have demonstrated the effectiveness of the Mamba architecture in single-channel speech enhancement. This paper aims to adapt SEMamba for multi-channel applications with only a small increase in parameters. The resulting system, MC-SEMamba, achieved results on the CHiME3 dataset that were comparable or even superior to several previous baseline models. Additionally, we found that increasing the number of microphones from 1 to 6 improved the speech enhancement performance of MC-SEMamba.

\end{abstract}

\begin{IEEEkeywords}
Mamba, Multi-channel speech enhancement
\end{IEEEkeywords}

\section{Introduction}
The goal of speech enhancement (SE) is to recover high-quality speech from distorted inputs, improving both its intelligibility and quality while facilitating various downstream tasks~\cite{loizou2007speech, vary2023digital}. Consequently, SE has become a crucial front-end process for a wide variety of speech-related applications. Recent advancements in deep learning (DL) have positioned neural networks (NN) as the foundational models for monaural SE tasks~\cite{lu2013speech, xu2014regression, LSTM_weninger2015speech, zhao2018convolutional}, as they consistently outperform traditional digital signal processing (DSP)-based methods~\cite{wang2018supervised}. This transition from DSP to DL is also apparent in multi-channel speech enhancement (MCSE), where DL-based methods such as those described in~\cite{araki2015exploring, wang2020complex,lu2022espnet, tan2022neural,yang2023mcnet} have been proposed as alternatives to conventional optimal beamformers, such as the Minimum Variance Distortionless Response (MVDR)~\cite{capon1969high, habets2010mvdr} filter. Nevertheless, a common criticism of numerous DL-based methods is their escalating requirement for additional parameters to increase performance. 


The earliest NNs for sequence modeling included the 
long short-term memory (LSTM)~\cite{hochreiter1997long} 
and gated recurrent units (GRU)~\cite{cho-etal-2014-learning}, which use nonlinear dynamics and mechanisms such as gating to model state transitions. 
However, these architectures generally underperform when the sequence length exceeds a certain threshold. This limitation 
has been addressed by the transformer architecture~\cite{NIPS2017_3f5ee243}, 
which is ubiquitous in large-scale sequence models. 
Despite this, concerns regarding the number of parameters and computation costs have emerged for applications bounded by stringent parameter limitations.
This prompted Gu et al.~\cite{gu2023mamba} to introduce an alternative sequence model for transformers, Mamba, which achieved comparable or even superior performance in sequence modeling across various tasks~\cite{li2024spmamba, jiang2024dual, yadav2024audio}.

Before the emergence of the Mamba architecture, transformer-based models such as convolution-augmented transformers (Conformers)~\cite{gulati2020conformer} served as standard architectures for several SE models~\cite{kim2021se, abdulatif2024cmgan, lu23e_interspeech}.
In~\cite{lu23e_interspeech}, Lu et al. proposed MP-SENet, which achieved outstanding 
single-channel SE performance on the VoiceBank+DEMAND~\cite{valentini2016investigating, thiemann2013diverse} 
public dataset. 
Similar to CMGAN~\cite{abdulatif2024cmgan}, 
MP-SENet 
utilizes the Conformer architecture to extract sequential information across the frequency and time axes. 
Inspired by the success of MP-SENet 
and Mamba
, Chao et al.~\cite{chao2024investigation} proposed SEMamba which explored the potential of 
substituting the Conformer blocks of MP-SENet with Mamba blocks. 
Although the parameter count of the most advanced version of SEMamba was not lower than that of MP-SENet, the results of SEMamba
, including performance metrics, indicated that the Mamba architecture is a viable alternative to Conformers in SE.

This study extends the SEMamba model to the MCSE framework by slightly adjusting the network configuration to enable it to 
learn spatial information of the signals received by 
a microphone array without significantly increasing the number of parameters. The resulting model, Multi-channel SEMamba (MC-SEMamba), is an efficient MCSE system that achieves SE performance comparable to or even exceeding those of several previous methods.
Prior to this study, Quan et al. of~\cite{10570301} proposed oSpatialNet, an online version of their previous work, SpatialNet~\cite{quan2024spatialnet}. The best version of oSpatialNet utilizes Mamba blocks to learn spatial and temporal features, validating the effectiveness of incorporating the Mamba architecture into the MCSE deep learning framework. 
However, 
the effectiveness of using Mamba in a non-casual MCSE framework, where Mamba is trained to learn bidirectional temporal and frequency information, was not fully investigated in~\cite{10570301}. 
In addition, we examine how the number of microphones affects the performance of the proposed MC-SEMamba system.



\section{Related Works}

\subsection{Mamba}\label{subseq:Mamba}
Despite their high-quality performance in sequence modeling, transformer-based neural networks entail
heavy computational demands and exhibit quadratic scaling complexity.
Consequently, Mamba was proposed to reduce the complexity of efficient sequence modeling while maintaining or even exceeding the performance of transformers. 
The use of previous sequence models such as LSTMs has 
suggested that the essential components of most sequential data should be modeled via nonlinear dynamics. In contrast, the Mamba architecture utilizes an input-dependent time-varying state space model (SSM), S6, to learn linear dynamics of the latent states as shown in Fig. \ref{fig:SSM_S6_diagram} and below: 

\begin{figure}
    \centering
    \includegraphics[width=1\linewidth]{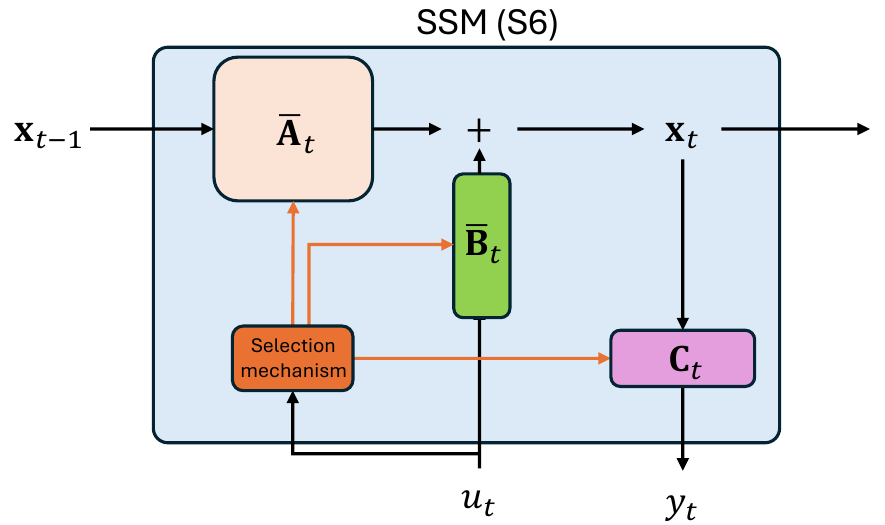}
    \caption{Diagram of S6 
    }
    \label{fig:SSM_S6_diagram}
    \vskip -0.1in
\end{figure}

\begin{equation}\label{eq:SSM_1}
\begin{aligned}
    \mathbf{x}_{t} &= \mathbf{\overline{A}}_t \, \mathbf{x}_{t-1} + \mathbf{\overline{B}}_t \, u_t \\
    y_{t} &= \mathbf{C}_t \, \mathbf{x}_{t}, 
\end{aligned}
\end{equation}
where $\mathbf{x}_{t} \in \mathbb{R}^N$, $u_t \in \mathbb{R}$, and $y_t \in \mathbb{R}$ represent the 
latent state, 
input, and output at time $t$ of the SSM. The matrices $\mathbf{\overline{A}}_t \in \mathbb{R}^{N \times N}$, $\mathbf{\overline{B}}_t \in \mathbb{R}^{N \times 1}$, $\mathbf{C}_t \in \mathbb{R}^{1 \times N}$ specify the parameters 
of the discretized SSM at time $t$. The S6 model utilizes a selection mechanism that updates $\mathbf{\overline{A}}_t$, $\mathbf{\overline{B}}_t$, and $\mathbf{C}_t$ based on selectively placing stronger emphasis on the more relevant contents of the input. 
Gu et al. of~\cite{gu2023mamba} suggested that nonlinear transformations be added before and after the S6 block, 
which corresponds to a combination of a gated multi-layer perceptron 
and the H3~\cite{fu2023hungry} architecture. 
In addition, Mamba is recognized for its hardware-specific algorithms to accelerate the training process, making it a suitable alternative for transformers for time-constrained applications. 

\subsection{SEMamba}\label{subseq:SEMamba}
Building on the success of MP-SENet in single-channel SE tasks, Chao et al. of~\cite{chao2024investigation} explored the effectiveness of replacing the Conformer blocks in MP-SENet with Mamba blocks. The results were comparable, if not better, than those of previous state-of-the-art SE models. 
Similar to MP-SENet, SEMamba takes the compressed magnitude spectrum and original phase spectrum of the noisy signal 
$(\mathbf{X}_\mathrm{cmag},\mathbf{X}_\mathrm{pha})$ as inputs, and outputs the compressed magnitude spectrum and the phase spectrum of the denoised signal $(\mathbf{Y}_\mathrm{cmag},\mathbf{Y}_\mathrm{pha})$, as shown below: 
\begin{equation}\label{eq:SEMamba}
\begin{aligned}
    f_{\text{SEMamba}}: \mathbb{R}^{T \times F} \times \mathbb{R}^{T \times F} &\to \mathbb{R}^{T \times F} \times \mathbb{R}^{T \times F}\\ 
    (\mathbf{X}_\mathrm{cmag},\mathbf{X}_\mathrm{pha}) &\mapsto (\mathbf{Y}_\mathrm{cmag},\mathbf{Y}_\mathrm{pha}), 
\end{aligned}
\end{equation}
where $T$ and $F$ are the number of frames and frequency bins of the spectrum, respectively. 
SEMamba follows the MetricGAN~\cite{fu2021metricgan+} framework, in which a speech assessment model acts as a discriminator, steering the generator toward improving speech signals according to a designated quality metric.
The generator network comprises two main stages. The first stage 
utilizes a dense encoder, where a Dilated Dense network~\cite{pandey2020densely} is employed 
to extract features at various resolutions. A sequence consisting of a 2D convolutional layer, instance normalization~\cite{ulyanov2016instance}, and parametric ReLU (PReLU)~\cite{he2015delving} is applied both before and after the Dilated DenseNet. 
The output of the dense encoder is subsequently passed through 
a sequence of blocks, each containing two bidirectional Mamba units that capture sequential information across time and frequency, respectively. Each Mamba unit is followed by a 1D transposed convolutional layer and a residual connection. 
This approach, which involves stacking sequence models to separately capture sequential information across either the time or frequency axes, 
has been widely adopted in prior studies~\cite{abdulatif2024cmgan, lu23e_interspeech, yang2023mcnet, quan2024spatialnet}. 

The second stage is the mask and phase decoder 
stage, in which the output from the preceding stage is processed using mask and phase decoders. As 
in the first stage, each decoder initially applies a Dilated DenseNet to the output from the previous stage. 
The phase decoder generates 
the denoised phase spectrum, whereas the mask decoder produces a mask that is multiplied element-wise by 
the compressed noisy magnitude spectrum, yielding the estimated compressed denoised magnitude spectrum. These can be used to calculate the terms required for the loss function.


\begin{figure*}[!h]
    \centering
    \includegraphics[width=0.92\linewidth]{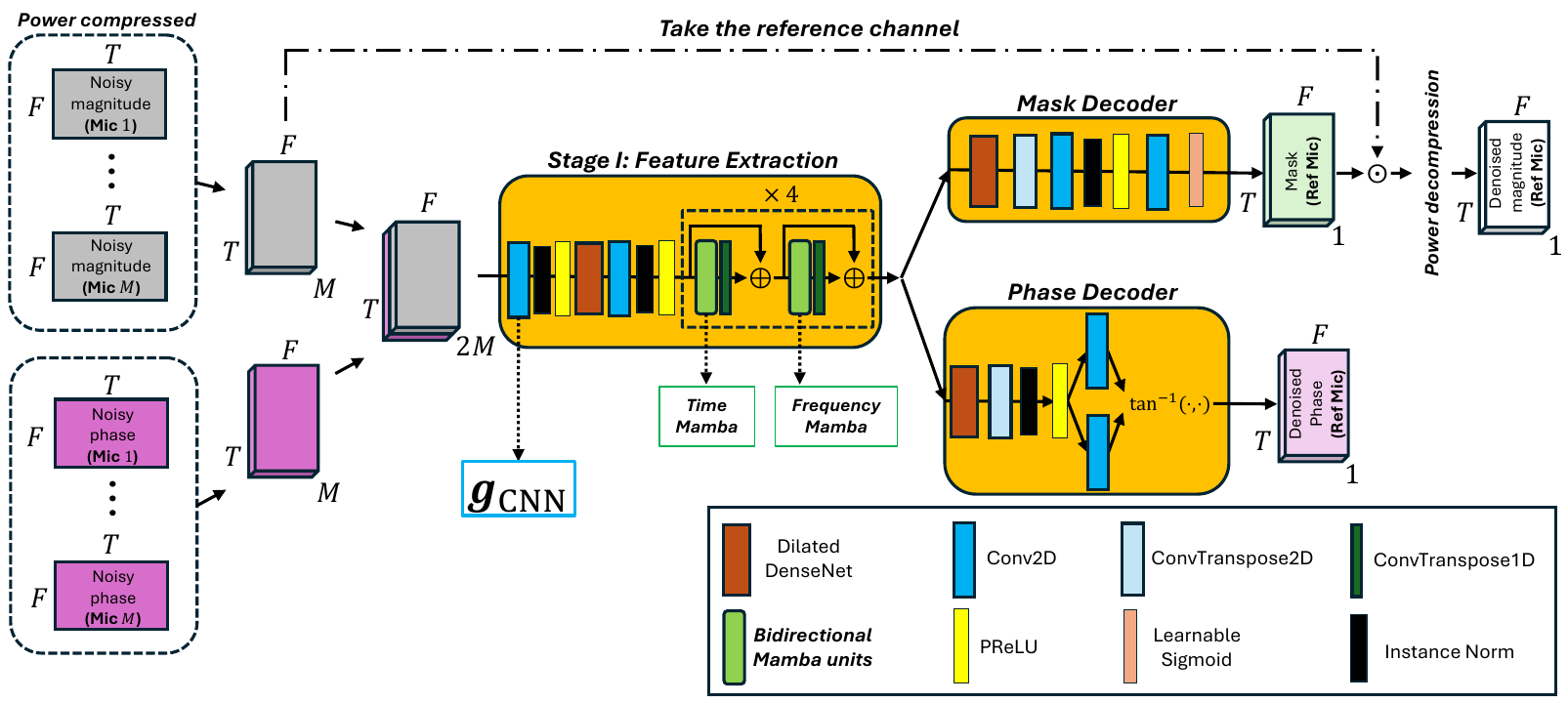}
    \caption{MC-SEMamba generator diagram. The architectural difference between SEMamba and MC-SEMamba is in $\boldsymbol{\mathit{g}}_\mathrm{CNN}$. Different blocks with the same color may have different types of parameters (e.g., kernel size). Operations such as tensor permutation are omitted for simplicity. The learnable sigmoid was proposed in~\cite{fu2021metricgan+}.}
    \label{fig:MC-SEMamba_generator_Diagram}
\end{figure*}

\section{MC-SEMamba}

A simple way 
to extend SEMamba for MCSE is to increase only the number of parameters at the front end of the model 
to account for recordings from the additional microphones. The resulting system, MC-SEMamba (Fig. \ref{fig:MC-SEMamba_generator_Diagram}), 
processes the compressed magnitude and original phase spectra of all microphone recordings, 
$\{(\mathbf{X}^{(m)}_\mathrm{cmag},\mathbf{X}^{(m)}_\mathrm{pha})\}^{M-1}_{m=0}$, where $M$ denotes the number of microphones, and outputs the compressed magnitude spectrum and the phase spectrum of the enhanced reference channel, $(\mathbf{Y}^\mathrm{ref}_\mathrm{cmag},\mathbf{Y}^\mathrm{ref}_\mathrm{pha})$, as shown below: 
\begin{equation}\label{eq:MCSEMamba}
\begin{aligned}
    f_{\text{MC-SEMamba}}: 
        \mathbb{R}^{M \times T \times F} \times \mathbb{R}^{M \times T \times F} &\to \mathbb{R}^{T \times F} \times \mathbb{R}^{T \times F}\\
    \big\{(\mathbf{X}^{(m)}_\mathrm{cmag},\mathbf{X}^{(m)}_\mathrm{pha})\big\}^{M-1}_{m=0} & \mapsto (\mathbf{Y}^\mathrm{ref}_\mathrm{cmag},\mathbf{Y}^\mathrm{ref}_\mathrm{pha}).
\end{aligned}
\end{equation}
Because the input layer of SEMamba is a 2D convolutional layer, denoted as $\boldsymbol{\mathit{g}}_\mathrm{CNN}$, adapting the model into MC-SEMamba involves increasing the number of input channels of the convolution layer, 
resulting in a very small 
increase in the number of parameters of the model. The primary goal is to achieve a noticeable improvement in  SE performance when transitioning from single-channel SE to multi-channel SE while incurring only a small 
increase in the model size.
\section{Experimental Results and Analysis}\label{sec:exp_results}
This section is organized into two parts. First, we compared the MCSE performance of a six-channel MC-SEMamba (MC-SEMamba-6) with that of existing methods. The second part provides a brief discussion on how 
changes in the number of microphones
affects the performance of MC-SEMamba. 
The SE scores used for comparison include the short-time objective intelligibility (STOI)~\cite{taal2011algorithm}, signal-to-distortion ratio (SDR)~\cite{vincent2006performance}, scale-invariant signal-to-distortion ratio (SI-SDR)~\cite{le2019sdr}, 
and 
both narrowband and wideband versions of the perceptual evaluation of speech quality (PESQ)~\cite{rix2001perceptual}. 
The raw narrowband MOS 
and narrowband MOS-LQO versions of PESQ are referred to as PESQ (N-1) and PESQ (N-2), respectively. The wideband PESQ is denoted as PESQ (W).

\subsection{Dataset}
ur experiments were performed using the CHiME3~\cite{barker2015third} simulated dataset, a widely recognized dataset used to train 
MCSE systems~\cite{yoshioka2015ntt, wang2020complex, yang2023mcnet, quan2024spatialnet}. The dataset contains 7138 training samples, 1640 validation samples, 1320 testing samples. Each sample is a six-channel recording generated from the software provided by Barker et al.~\cite{barker2015third},  designed to simulate a person speaking in a noisy environment while holding a tablet equipped with six microphones arranged in 
a rectangular pattern. 
The noisy environments include ``Cafe", ``Street", ``Bus" and ``Pedestrian". 
The waveforms were sampled at 16,000 Hz and 
converted into magnitude and phase spectra representations using a Hann 
window with a size of 400, a hop size of 100, and a 400-point FFT. 
The magnitude spectrum was compressed with a power compression factor of 0.3. 
This compression scheme was shown in~\cite{li2022glance} to enhance SE performance. 
\subsection{Network configurations}
Here we discuss the parameters of $\boldsymbol{\mathit{g}}_\mathrm{CNN}$, which vary according to the number of microphone inputs, $M$. As we increase $M$ by one, the number of parameters increases by only 128. Specifically, the number of parameters in $\boldsymbol{\mathit{g}}_\mathrm{CNN}$ is specified by the 
number of input channels $c_\mathrm{in}$, output channels $c_\mathrm{out}$, and kernel size $k$. The default values of $c_\mathrm{out}$ and $k$ are 64 and 1, 
respectively. As we increase the number of microphones by one, we increase $c_\mathrm{in}$ by two which corresponds to the magnitude and phase spectra of the additional microphone's data. Thus, the total number 
of parameters introduced for one additional microphone input is $128=2 \times 1 \times 64$. The remaining network configurations, including those of the discriminator, were directly referenced from~\cite{lu23e_interspeech, chao2024investigation}\footnote{https://github.com/RoyChao19477/SEMamba}.

\subsection{Training Specifics}
Among the six available microphone recordings in CHiME3, 
the fifth microphone was used as the reference channel, as in~\cite{wang2020complex, yang2023mcnet} and~\cite{quan2024spatialnet}. The software provided by Barker et al.~\cite{barker2015third} allows us to obtain the clean signal for the reference channel. We then used this as the clean reference signal to calculate the SE scores\footnote{We suspect that the discrepancy in several scores (especially the SDR) of the noisy reference channel is partially due to a slight difference in the choice of channel for the clean reference signal.}. We used AdamW~\cite{loshchilov2018decoupled} as the optimizer for both the generator and discriminator, with a learning rate of $5 \times 10^{-4}$, and $\beta$ parameters: $(\beta_1, \beta_2)=(0.8, 0.99)$. The learning rate had an exponential decay factor 
of 0.99 per epoch. We trained MC-SEMamba for 50 epochs with a batch size of one 
and selected the 
epoch with the highest PESQ score on the validation set. 
\subsection{Experimental Analysis}
\paragraph{\textbf{Comparison With Previous Methods}}
We compared our method with the small and large versions of SpatialNet (SpatialNet-s and SpatialNet-L)~\cite{quan2024spatialnet}
and several of their baselines, including the Multi-cue Net (McNet)~\cite{yang2023mcnet},  FT-JNF~\cite{tesch2022insights}, the Narrow-band deep filter~\cite{li2019narrow} (NBDF)
, Channel-Attention Dense U-Net~\cite{tolooshams2020channel} (CA Dense U-Net),
the oracle MVDR beamformer, 
the multichannel nonnegative matrix factorization beamformer~\cite{shimada2019unsupervised} (MNMF-BF), and the complex spectral mapping method (CSM) by Wang et al.\cite{wang2020complex}. 

The results are presented in 
Table \ref{tab:compare_wth_others_chime3_added_SpatialNet}. We also compared the number of parameters of our model with those of several other methods, as shown in Table \ref{tab:n_params_compare_partial}. Because the discriminator of MC-SEMamba is not used during inference, the MC-SEMamba-6 parameter count in Table \ref{tab:n_params_compare_partial} refers only to that of the generator. All results in Tables \ref{tab:compare_wth_others_chime3_added_SpatialNet} and \ref{tab:n_params_compare_partial} except those from MC-SEMamba have been quoted from either the original papers or from the McNet and SpatialNet papers. 
Table \ref{tab:compare_wth_others_chime3_added_SpatialNet} shows that 
the PESQ (N-2), PESQ (W), and STOI scores of MC-SEMamba-6 are higher than those of SpatialNet-s and SpatialNet-L. 
Although the CSM method yields the highest STOI score, our method uses significantly fewer parameters\footnote{The CSM method that produced the scores shown in Table \ref{tab:compare_wth_others_chime3_added_SpatialNet} required two NNs, each with a parameter count of about 13 million.}, as detailed in Table \ref{tab:n_params_compare_partial}.

\begin{table}[!t]
\caption{
Average SE scores for ``Noisy'', MC-SEMamba, and other methods on CHiME3 (*, $\dag$  means quoted from the McNet/SpatialNet, and other original papers, respectively.)}\label{tab:compare_wth_others_chime3_added_SpatialNet}
\setlength{\tabcolsep}{3.1pt} 
\centerline{
\begin{tabular}{lcccccc}
\toprule
    & \textbf{PESQ} & \textbf{PESQ} & \textbf{PESQ} & \textbf{STOI} & \textbf{SI-SDR} & \textbf{SDR} \\
    & (N-1)       & (N-2)   &(W) &(\%)  & & \\
\cmidrule(r){1-7}
    
    Noisy (CH5)                                 & 2.18     & 1.82      & 1.27    & 87.0     & 7.5    & 7.5    \\
    MNMF-BF\cite{shimada2019unsupervised} $\dag$                              & 2.70     & -         &  -      & 94.0     & -      & 16.2   \\
    Oracle MVDR *                               & -        & 2.49      & 1.94    & 97.0     &  17.3  & 17.7  \\
    CA Dense U-Net\cite{tolooshams2020channel} $\dag$ & -        & -         & 2.44    & -        & -      & 18.6  \\
    NBDF\cite{li2019narrow} *                 & -  & 2.74      & 2.13      & 95.0      & 15.7   & 16.6    \\
    FT-JNF\cite{tesch2022insights} ~\cite{quan2024spatialnet} * & - & 3.20     & 2.61     & 96.7     & 17.6    & 18.3 \\
    McNet\cite{yang2023mcnet} *              & -    & 3.38      & 2.73      & 97.6      & 19.2   & 19.6    \\
    CSM~\cite{wang2020complex} $\dag$        & \bf{3.68}    & -      &  -      & \bf{98.6} & 22.0   & \bf{22.4} \\
    SpatialNet-s~\cite{quan2024spatialnet} *  &  -                      & 3.49      &  2.88     &  98.3     & \bf{22.1}   & 22.3    \\
    SpatialNet-L~\cite{quan2024spatialnet} *  &  -                      & 3.45      &  2.89     &  98.2     & 21.8   & 22.1   \\
    MC-SEMamba-6                             &  3.48   & \bf{3.50}      & \bf{3.07} &  98.5     & 21.3   & 21.9 \\
\bottomrule
\end{tabular}}
\end{table}

\paragraph{\textbf{Impact of Choice of Microphones}}
We also conducted experiments to compare the performance of MC-SEMamba with various numbers of microphone inputs and 
tested six models, each with a unique number of microphones ranging from 1 to 6. 
Our choice of microphones for each model is detailed as follows: 
The model trained using a single microphone (denoted as 1 Mic) used the reference microphone (i.e., the $5^{th}$ channel, CH5).  
For two and three mics, we maintained the linear array structure by choosing the neighboring mics, which correspond to (CH4, 5) and (CH4, 5, 6), respectively. For the models corresponding to 4 and 5 mics, the choice of microphones was (CH1, 4, 5, 6) and (CH1, 2, 4, 5, 6), respectively. 
The weights for each model were selected individually from the epoch with the highest PESQ 
score on the validation set within 50 epochs. 

The results are summarized in Table 
\ref{tab:compare_wth_various_nmics_chime3}. 
Within this set of experiments, the model that used all six microphones yielded the best results, with a noticeable difference compared with single-channel SEMamba. We also note that the PESQ (W) and STOI results from using five microphones are higher than those of both versions of SpatialNet, which suggests that five microphones are sufficient for MC-SEMamba to enhance the quality and intelligibility of noisy speech signals for the CHiME3 dataset. 

\begin{table}[!t]
\caption{Number of parameters in different models (M $ = 10^6$).
} \label{tab:compare_wth_various_nmics_chime3} 
\setlength{\tabcolsep}{3.5pt} 
\centerline{
\begin{tabular}{cccccc}
\toprule
    & \textbf{McNet}* & \textbf{CSM}$\dag$ & \textbf{SpatialNet-s}* & \textbf{SpatialNet-L}* & \textbf{MC-SEMamba-6} \\
    & & & (16 kHz) &(16 kHz) &  \\
      & 1.9 M & 26.0 M & 1.6 M & 7.3 M & 2.3 M\\
\bottomrule
\end{tabular}} \label{tab:n_params_compare_partial}
\end{table}

\begin{table}[!t]
\caption{Average SE scores for MC-SEMamba using various mics on the CHiME3 test set
: 1 Mic (CH5), 2 Mics (CH4, 5), 3 Mics (CH4, 5, 6), 4 Mics (CH1, 4, 5, 6), 5 Mics (CH1, 2, 4, 5, 6)}\label{tab:compare_wth_various_nmics_chime3} 
\centering
\begin{tabular}{l  c  c  c  c  c c}
\toprule
    & \textbf{PESQ} & \textbf{PESQ} & \textbf{PESQ} & \textbf{STOI} & \textbf{SI-SDR} & \textbf{SDR} \\
    & (N-1) & (N-2)       & (W)   &(\%)  &\\
\cmidrule(r){1-7}
    Noisy        & 2.18 & 1.82 & 1.27 & 87.0 & 7.5 & 7.5   \\
    1 Mic        & 3.18 & 3.09 & 2.43 & 95.3 & 16.1 & 16.6  \\
    2 Mics       & 3.34 & 3.31 & 2.73 & 97.0 & 17.7 & 18.2  \\
    3 Mics       & 3.34 & 3.32 & 2.76 & 97.6 & 19.7 & 20.1  \\
    4 Mics       & 3.33 & 3.31 & 2.77 & 97.7 & 19.5 & 20.1  \\
    5 Mics       & 3.45 & 3.47 & 3.06 & 98.4 & 20.6 & 21.6  \\
    6 Mics       & \bf{3.48} & \bf{3.50} & \bf{3.07} & \bf{98.5} & \bf{21.3} & \bf{21.9}  \\
\bottomrule
\end{tabular}
\label{table:1NN error}
\end{table}

\section{Conclusion and Future Work}\label{sec:conclusion}
We proposed MC-SEMamba, a simple extension of SEMamba that can learn the spatial information of speech recordings from microphone arrays. The results from our experiments are comparable to, or even exceed, 
those of several 
state-of-the-art MCSE methods, with a noticeably smaller parameter count than that of the CSM method. We also briefly discussed the effect of increasing the number of microphones on the MCSE performance of the proposed method.
In the future, we plan to further investigate how different array geometries can potentially affect the MCSE performance of MC-SEMamba. 



\bibliographystyle{IEEEtran}
\bibliography{mcsemamba}

\end{document}